\documentclass[conference]{IEEEtran}

\usepackage{cite}
\usepackage{amsmath,amssymb,amsfonts}
\usepackage{algorithmic}
\usepackage{tikz}
\usetikzlibrary{positioning, fit, calc}
\usepackage{hyperref}
\usepackage{graphicx}
\usepackage{textcomp}
\usepackage{multirow}
\usepackage{float}
\usepackage{booktabs}
\usepackage{tabularx}
\usepackage[utf8]{inputenc}
\usepackage[T1]{fontenc}
\usepackage[most]{tcolorbox}
\usepackage[margin=1in]{geometry}

\usepackage{tcolorbox}
\usepackage{xcolor}
\usepackage{enumitem}

\usepackage[table,xcdraw]{xcolor}
\usepackage[normalem]{ulem}
\useunder{\uline}{\ul}{}

\newif\ifshowcomments 

\def\BibTeX{{\rm B\kern-.05em{\sc i\kern-.025em b}\kern-.08em
    T\kern-.1667em\lower.7ex\hbox{E}\kern-.125emX}}
\begin{document}

\title{Defeater Cards: Characterizing and Managing Safety Assurance Case Defeaters\\

}

\author{
  \begin{tabular}{c@{\extracolsep{0.5cm}}cc}
    Usman Gohar & Michael C. Hunter & Salil Purandare \\
    \textit{Dept. of Computer Science} & \textit{Dept. of Computer Science} & \textit{Dept. of Computer Science} \\
    \textit{Iowa State University} & \textit{Iowa State University} & \textit{Iowa State University} \\
    Ames, Iowa & Ames, Iowa & Ames, Iowa \\
    \\ 
    
    Jordan J. Rios & Myra B. Cohen & Robyn R. Lutz \\
    \textit{Dept. of Computer Science} & \textit{Dept. of Computer Science} & \textit{Dept. of Computer Science} \\
    \textit{Iowa State University} & \textit{Iowa State University} & \textit{Iowa State University} \\
    Ames, Iowa & Ames, Iowa & Ames, Iowa \\
  \end{tabular}
}

\maketitle

\begin{abstract}

Safety assurance cases provide structured justifications that safety-critical systems meet their safety requirements. Recently, the notion of defeaters has emerged as a rigorous means of challenging the validity of safety arguments. Examples of defeaters might include overly strict claims, unreliable evidence, or reasoning gaps. However, defeaters remain ad hoc, lack structured support for critical reflection, are inconsistently described, are difficult to review, and lack documentation standards. To address this, we propose Defeater Cards, a new structured documentation artifact for systematically characterizing, reasoning about, and managing defeaters in safety cases. Drawing on a literature survey and thematic analysis, we identify documentation criteria that inform the card's structure, based on the 5W1H framework. Defeater Cards are designed to support informed analysis and evolution, improve traceability and auditability, and enable the reuse of defeater knowledge
across systems and product variants.  We demonstrate their applicability through two cross-domain case studies, showing how they expose hidden assumptions, surface reasoning gaps, and support ongoing safety assurance case evolution. To support adoption and community reuse, we also release an open-source repository of defeater cards as a baseline upon which researchers and practitioners can build and describe lessons learned.

\end{abstract}

\pagestyle{plain}
\section{Introduction} 
\label{sec:introduction}

Demonstrating the safety and reliability of safety-critical cyber-physical systems (CPS), such as autonomous vehicles and medical devices, is often a prerequisite for their deployment \cite{johnson1998178b,palin2011iso, koopman2019safety}. Safety assurance cases support this process by providing structured arguments that link safety requirements to supporting evidence, justifying that a system can perform safely in its intended environment \cite{ACWG21,bloomfield2020assurance, Knight12}. They serve both as internal artifacts to guide development and as formal justifications reviewed by regulators for certification \cite{maksimov2019survey, denney2012advocate}.  Over the past decades, extensive research has supported the construction of assurance cases, through formal notations such as Goal Structuring Notation (GSN) \cite{kelly2004goal} and through the release of various tools \cite{maksimov2019survey,10311213,denney2012advocate}.

In practice, however, assurance cases are susceptible to obstacles \cite{jLamLet00} to the soundness of their claims that the safety goals of the deployed product are adequately satisfied. These obstacles, termed \emph{defeaters}, are any factors, conditions, or events that weaken or invalidate the safety claims made about the system \cite{denney2011towards, goodenough2015eliminative, gohar2025taxonomy, LetierL25}. Defeaters arise from infeasible safety requirements, incorrect assumptions, incomplete evidence, contextual gaps, or unforeseen system interactions. If left unaddressed, they compromise the trustworthiness of the assurance case, leading to unwarranted overconfidence in system safety and resulting in failures~\cite{cave2006independent}. For example, consider an assurance case for a small Uncrewed Aircraft System (sUAS) with the requirement, “The sUAS can safely fly in current wind conditions”. This could be challenged by the defeater: “Unless unexpected wind shear or turbulence occurs at higher altitudes” \cite{hunter2025safecert}. 

To date, prior work has focused primarily on identifying and mitigating defeaters during initial assurance case development through semantic analysis, formal reasoning \cite{murugesan2023semantic,rushby2015understanding,muram2023attest,yuan2016automatically}, and human-in-the-loop LLM approaches \cite{gohar2024codefeater,AISupported}. However, defeaters emerge throughout the safety case lifecycle, including during regulatory audits and ongoing maintenance. In particular, audits play an essential role in (1) surfacing overlooked argument gaps, (2) validating assumptions, and (3) ensuring that emerging vulnerabilities are addressed before deployment \cite{konigstorfer2022ai}. The effectiveness of these stages heavily depends on process transparency, access to contextual information, and the rationale behind the defeaters. This aligns with our experience in our work on safety-critical systems, where project risk and cost grow when defeaters are inadequately documented, and the system or its operational context evolves \cite{Lutz07, LutzLLKHMS12, Lutz22}. 

Yet, current practices for documenting defeaters remain limited and fragmented across notations: processes are often ad hoc \cite{maksimov2019survey,gohar2025taxonomy}, and defeaters are inconsistently tracked \cite{ACWG21} and poorly documented \cite{bloomfield2021safety, rushby2015understanding}. Recently, there have also been growing concerns about the rigor and validity of the defeater process~\cite{ACWG21,assurance2021assurance}, with research emphasizing that documentation must ensure reproducibility and auditability to enable critical review \cite{puhlfurss2025model,bloomfield2020assurance}. However, these 
principles are inadequately supported by current defeater documentation. Despite repeated calls, the lack of consensus on documentation best practices across notations further impedes adoption and standardization, a challenge echoed in adjacent domains for evaluating systems (e.g., AI safety \cite{feffer2024red, Anthropic}). As cyber-physical systems increasingly integrate AI and operate in dynamic environments (e.g., drones), where assumptions and risks require continuous reevaluation \cite{denney2015dynamic}, a systematic approach to identifying, reasoning about, and documenting defeaters becomes essential for safety management, certification, and maintenance.

As a step toward that, 
we propose \emph{Defeater Cards}: a lightweight, standardized framework for documenting defeaters across six evaluation dimensions, based on a literature review of current challenges and best practices. Inspired by documentation strategies in AI (Data Cards \cite{pushkarna2022data}, Model Cards \cite{mitchell2019model}) and Requirements Engineering (RE) (Snow Cards \cite{robertson2000volere}), Defeater Cards explicitly capture contextual information and design rationale, 
support comprehensive analysis, informed review, and reuse across product variants, properties unsupported by current fragmented methods. Our approach is particularly well-suited to emerging modular certification efforts, such as certificates for sUAS, where safety assurance must be distributed across components (e.g., drone type, pilot, flying conditions) and reused in different contexts \cite{hunter2025safecert,hunter2024family}. To our knowledge, this is the first work addressing this gap in defeater documentation that currently impedes safety-case audits and long-term maintenance.

Overall, this work makes four key contributions: 

\begin{itemize}
    \item We identify 16 major documentation gaps and challenges, formalized as criteria, in current practice through a structured literature survey.
    \item We propose an audit-friendly standardized artifact, the Defeater Card, to characterize and reason about defeaters.
    \item Initial validation of our framework for two CPS case studies, and develop a dataset of defeaters and integrate them with an existing taxonomy to support use by researchers.
\end{itemize}

The rest of the paper is organized as follows: In Section \ref{sec:background}, we provide motivation and related work. Section \ref{sec:methodology} outlines our design methodology and the results of our survey, while Section \ref{sec:card} introduces Defeater Cards. Section \ref{sec:applications} discusses our two case studies. Finally, we discuss the implications of our work in Section \ref{sec:discussion} and Section \ref{sec:conclusion} gives concluding remarks. \\

\section{Motivation \& Related Work}
\label{sec:background}

\begin{figure}[t!]
    \centering
    \includegraphics[scale = 0.70]{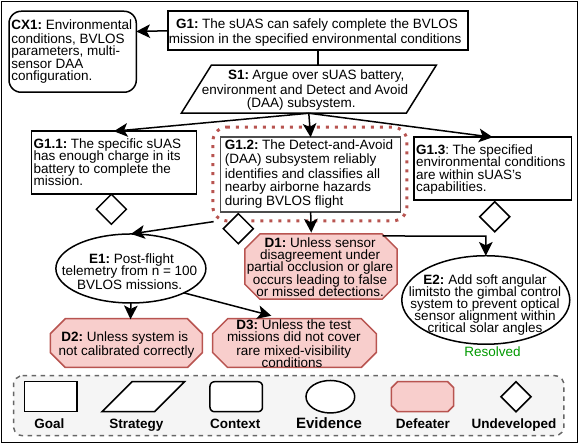}
    \caption{sUAS Safety Assurance Case fragment with example defeaters} 
    \label{fig:AC}
    \vspace{-2.5mm}
\end{figure}

\subsection{Current Practices and Challenges}

Defeaters represent conditions, assumptions, or evidence gaps that challenge the validity of safety claims in an assurance case, playing a critical role in evaluating safety-critical systems \cite{denney2011towards,goodenough2015eliminative}
Unlike Fault Tree Analysis, which identifies events that can contribute to hazards, or FMECA, which evaluates failure consequences, defeaters specifically target potential weaknesses in assurance arguments \cite{Knight12}. 
Figure~\ref{fig:AC} shows a fragment of an sUAS assurance case that we developed with example defeaters (red boxes). This fragment relates to new US Federal Aviation Administration regulations for flying Beyond Visual Line of Sight (BVLOS). 

While defeater analysis strengthens the robustness of assurance cases, its practical application remains limited by the lack of systematic methods for documenting how defeaters are identified, evaluated, and resolved \cite{bloomfield2021safety,bloomfield2020assurance}. The lack of transparency creates further risks, particularly the potential to game the audits and project false confidence in the assurance case. One barrier to meaningful transparency is a lack of standards for reporting \cite{staufer2025audit,courtois1993documentation}, as the current landscape of defeater analysis is largely ad hoc and disjointed across different notations and frameworks \cite{gohar2025taxonomy,rushby2015interpretation}. Finally, defeaters typically emerge through dialectic reasoning among developers, auditors, and domain experts, yet current documentation provides little structured support for capturing the rationale and contextual information needed. This often makes it hard to review and audit defeater-based safety assurance cases.

These challenges are amplified by a shift from static, controlled systems \cite{denney2015dynamic} to dynamic, autonomous, AI-driven systems operating in uncertain environments, where assurance cases may require  ongoing updates, e.g., as new field data emerges \cite{koopman2019safety,gohar2024codefeater}. Current approaches provide inadequate support for ongoing audit and maintenance \cite{assurance2021assurance,ACWG21}, leaving reviewers without critical metadata on system capabilities, limitations, and the rationale for assumptions. Therefore, a more systematic and standardized approach to identifying and documenting defeaters can help practitioners anticipate issues throughout the product lifecycle. Defeater Cards attempt to close this gap by unifying notation and reporting key background information, design choices, and justifications. 

\subsection{Related Work}

\noindent \textbf{Documentation Frameworks.} Structured documentation templates have supported transparency and reproducibility to address similar challenges in requirements engineering and adjacent domains. For instance, Snow Cards \cite{robertson2000volere} standardize requirements elicitation, while Model Cards \cite{mitchell2019model} and Data Cards \cite{pushkarna2022data} document the limitations and risks of AI models and the characteristics of datasets, respectively. In this work, we build on this line of efforts to introduce Defeater Cards for safety assurance cases.

\noindent \textbf{Evaluating and Assessing Assurance cases.} Prior research has proposed tools \cite{maksimov2019survey,denney2018tool,
di2020mmint} and methods \cite{cyra2011support,mayo2006structured,
panesar2010characterizing} to assess assurance case validity, typically focusing on structural soundness \cite{chowdhury2020systematic} or quantifying confidence and uncertainty \cite{duan2017reasoning,idmessaoud2024confidence,goodenough2012toward}. 
Other works explore whether LLMs could help review assurance cases  \cite{Graydon25, Yu25}. More recently, defeaters have gained traction as a complementary approach. Several works integrate defeaters into assurance frameworks \cite{bloomfield2024defeaters,
diemert2023incremental} or use semantic analysis \cite{murugesan2023semantic,muram2023attest,yuan2016automatically,rushby2015understanding} and LLMs \cite{gohar2024codefeater,AISupported} to identify logical inconsistencies and latent challenges. However, these efforts focus primarily on identification during the initial assurance case development, with limited attention to documentation practices that support audit and maintenance throughout the lifecycle.

Modern safety-critical systems, particularly those that are dynamic and AI-driven, require continuous reevaluation of assurance cases \cite{agrawal2019leveraging,javed2021towards,denney2015dynamic}.  
For instance, new evidence, shifting contexts, or invalidated assumptions can surface new defeaters or alter existing ones. Knowledge turnover and domain expertise loss compound 
these challenges, making it difficult to revisit earlier defeater reasoning. While continuous assurance frameworks \cite{schleiss2022towards,jaradat2015facilitating} monitor systems to trigger case updates, these require pre-specified triggers and do not support contextual changes \cite{carlan2021safety}. In this work, we argue that defeater-based assurance should be viewed through the lens of lifecycle maintenance and audit, with standardized documentation as a critical first step. 
\section{Defeater Cards Framework Design}
\label{sec:methodology}

To inform the structure and content of Defeater Cards, we first systematically identify documentation gaps and challenges through a literature review, distill them into high-level documentation criteria, and then operationalize these principles via the 5W1H framework. We primarily investigate two research questions:

\begin{figure}[t!]
    \centering
    \includegraphics[width = \columnwidth]{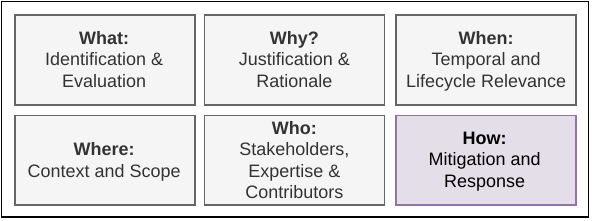}
    \vspace{-4mm}
    \caption{Structural overview of the Defeater Card: Categorizing identification (5Ws) and mitigation (1H) through the 5W1H framework \cite{waisbord20215ws}.}
    \label{fig:overview}
\end{figure}

\begin{itemize}[]
    \item \textbf{RQ1:} What defeater documentation gaps and best practices exist in safety assurance cases?

    \item \textbf{RQ2:} Can the Defeater Cards framework be applied across diverse safety-critical domains?
\end{itemize}

We follow the established Design Science Research methodology \cite{ceci2024defining,peffers2007design}: \textit{(1) Problem identification} (Section~\ref{sec:background}), \textit{(2) Objectives} through identifying gaps (RQ1), \textit{(3) Design \& development} by deriving criteria, and creating Defeater Cards, and \textit{(4-5) Demonstration and 
Evaluation through case studies (RQ2).}

\subsection{Methodology. } 

\begin{table*}[t]
\centering
\caption{Taxonomy of Defeater Documentation Criteria derived from literature
and cross-domain best practices, organized by theme and mapped to Defeater Card dimensions.}
\label{tab:requirements_taxonomy}
\resizebox{\textwidth}{!}{%
\begin{tabular}{lll@{}}
\toprule
\textbf{Documentation Criteria} 
& \textbf{Representative Sources} 
& \textbf{Defeater Card} \\
\midrule

\multicolumn{3}{l}{\textbf{Provenance}} \\

(R1) Documents the defeater's discovery source and methodology (e.g., red-teaming, formal analysis)
& \cite{gohar2025taxonomy, feffer2024red, ACWG21}
& What \\

(R2) Records the triggering event or condition that surfaced the defeater
& \cite{denney2015dynamic, hobbs2024driving}
& What + When \\

(R3) Links the defeater to the specific version and lifecycle phase in which it was identified
& \cite{muram2023attest, mitchell2019model}
& Metadata + When \\

\midrule
\multicolumn{3}{l}{\textbf{Auditability}} \\

(R4) Documents the background and expertise of  contributors (human or automated)
& \cite{feffer2024red, costanza2022audits, gebru2021datasheets} 

& Who \\
(R5) Maintains audit trails of contributor inputs and challenges
& \cite{ACWG21, puhlfurss2025model}
& Metadata \\

\midrule
\multicolumn{3}{l}{\textbf{Justification \& Scope}} \\

(R6) Provides rationale for identifying, prioritizing, or dismissing a potential defeater
& \cite{burge2006rationale, ACWG21, rushby2015understanding}
& Why \\

(R7) States underlying logical premises and assumptions supporting the defeater claim
& \cite{bloomfield2024defeaters, rushby2015understanding, mitchell2019model}
& Why \\
(R8) Assesses and document severity, likelihood, and safety-critical implications
& \cite{ACWG21, hobbs2024driving, feffer2024red}
& Why \\

(R9) Defines the operational context and environmental boundaries in which the defeater applies
& \cite{rushby2015understanding, bloomfield2021safety, mitchell2019model}
& Where \\

(R10) Specifies the system component or subsystem affected by the defeater
& \cite{bloomfield2024defeaters, bloomfield2021safety, hobbs2024driving}
& Where \\

\midrule
\multicolumn{3}{l}{\textbf{Mitigation Rigor \& Robustness}} \\

(R11) Specifies mitigation strategies with boundary conditions and limitations
& \cite{ACWG21, staufer2025audit, konigstorfer2022ai}
& How \\

(R12) Documents monitoring arrangements and re-evaluation triggers for unresolved or evolving defeaters
& \cite{denney2015dynamic, carlan2021safety,agrawal2019leveraging}
& How \\

(R13) Records residual risks remaining after mitigation
& \cite{hobbs2024driving, diemert2023incremental, bloomfield2021safety}
& How \\

(R14) Reports the reliability and validity of proposed mitigations and supporting evidence
& \cite{staufer2025audit, bloomfield2021safety, holloway2018explicate}
& How \\

\midrule
\multicolumn{3}{l}{\textbf{Dialectic Reasoning}} \\

(R15) Preserves challenge-response threads, rebuttals, and counterfactual scenarios
& \cite{bloomfield2024defeaters, ACWG21, goodenough2015eliminative }
& Why + Comments \\

(R16) Tracks the resolution status and decision-making process across review iterations
& \cite{ACWG21, hobbs2024driving, rushby2015interpretation} 
& How + Comments \\

\bottomrule
\end{tabular}%
}
\end{table*}

\noindent \textbf{Identifying Relevant Studies:
}
We employed 
systematic review methodologies \cite{keele2007guidelines,gohar2025taxonomy} to identify documentation challenges and gaps. We searched Google Scholar, ACM Digital Library, IEEE Xplore, and arXiv using the search string: ("assurance case" OR "safety case") AND ("defeater" OR "defeasible reasoning" OR "counter-argument" OR "assurance case weakeners" OR "deficits"), 
prioritizing papers on documentation, evaluation, or maintenance practices. We also performed backward and forward snowballing on seminal papers \cite{bloomfield2020assurance, rushby2015interpretation, ACWG21} to identify additional relevant work. To broaden our perspective, we also included a subset of documentation practices from AI evaluation and auditing \cite{staufer2025audit,feffer2024red}, software engineering design rationale \cite{burge2006rationale,burge2006rationale}, and RE (Snow Cards \cite{robertson2012mastering}) that share conceptual parallels. Finally, we drew on our team's experience building defeasible assurance cases. This process yielded $N=46$ papers, which are available in the artifact. 

\noindent \textbf{Thematic Analysis:} We analyzed the literature using thematic analysis \cite{Clarke2014} and open coding \cite{gohar2025taxonomy, vaismoradi2013content}. Two authors independently coded the data and reconciled discrepancies to establish a shared set of themes. Using an expert-based affinity diagramming approach, we synthesized these themes with our prior experience building safety cases to translate abstract codes into formal documentation criteria. These criteria were then mapped to the 5W1H dimensions \cite{waisbord20215ws} (shown in Figure \ref{fig:overview}) and iteratively refined through application to two example systems. \\

\noindent \textbf{Threats to Validity:} There are several threats to validity. First, the documentation criteria vary in granularity. Some are specific, reflecting well-established practices, while others are broader to allow flexibility across domains, contexts, and use cases. Nevertheless, the criteria is intended to inform relevant design choices rather than to provide an exhaustive list. Second, the literature on defeater documentation practices is still nascent and limited, which constrains the size of the study. We mitigate this by incorporating established principles from adjacent domains and frameworks, grounding criteria in empirically validated practices. Third, the Defeater Card reflects current documentation gaps and practices, as well as our subjective analysis of challenges. Future evaluations of defeater cards and new safety assurance standards may surface considerations not yet anticipated. The modular and flexible design is intended to support extensions. Finally, due to space limitations, only two case studies are discussed; broader empirical validation and user studies remain future work. We view the case studies as demonstrating initial feasibility and applicability, similar to prior work~\cite{mitchell2019model,ceci2024defining}.

\subsection{RQ1: Documented Criteria, Gaps and Best Practices} 

Our analysis identified five central themes that serve as documentation criteria for Defeater Cards, summarized in Table~\ref{tab:requirements_taxonomy}. These can also be interpreted as \textit{normative requirements} for Defeater Cards. Each of these has been mapped to the specific defeater card dimension.

\subsubsection{Provenance \& Traceability} The credibility of a defeater depends on knowing precisely where it came from and how it was identified. Provenance and traceability establish this foundation by linking each defeater to its discovery process, triggering conditions, and the relevant system context. The discovery source and methodology (\textbf{\textit{R1}}) indicate whether the identification was systematic or opportunistic, affecting completeness and reproducibility~\cite{gohar2025taxonomy, feffer2024red}. Recording the triggering event or condition (\textbf{\textit{R2}}) provides the rationale for why the defeater surfaced at a specific point~\cite{denney2015dynamic, ACWG21}. Linking the defeater to the version and lifecycle phase (\textbf{\textit{R3}}) preserves traceability as the system evolves~\cite{mitchell2019model, denney2015dynamic}. Together, these help prevent the loss of critical audit and reasoning information.

\subsubsection{Auditability} Existing works on defeaters predominantly focus on practitioners and initial defeater analysis of safety cases. An overlooked use case for defeaters is their use as an approach to review or audit an already developed safety case. However, current documentation practices do not adequately support auditability. This has been repeatedly reported as a critical shortcoming of current defeater analysis and documentation frameworks \cite{ACWG21}. Similar concerns have been raised in analogous processes such as red-teaming~\cite{feffer2024red} and AI documentation frameworks \cite{staufer2025audit,puhlfurss2025model}. In contrast to provenance, auditability concerns the process surrounding it, who conducted it, their expertise \textbf{\textit{(R4)}}, and whether the defeater process can be independently reviewed and verified \textbf{\textit{(R5)}}. This directly affects the credibility of the raised defeaters, as well as providing important signals of coverage and rigor \cite{gohar2025taxonomy}. For instance, model documentation literature~\cite{puhlfurss2025model, mitchell2019model} identifies contributor expertise as critical for surfacing potential bias in public-facing systems, a concern echoed in AI fairness research~\cite{costanza2022audits, gebru2021datasheets}. Therefore, standardized defeater documentation should provide enough contextual information to support review and audit.

\subsubsection{Justification \& Scope} A key feature of any robust and credible evaluation is providing justifications for any design and methodological choices. However, current practices in defeater analysis often leaves these justifications implicit, relying only on subjective interpretation. To ensure logical rigor and prevent frivolous challenges, it is imperative that defeaters are accompanied by a clear rationale. This justification should be grounded in scientific evidence, such as the experimental proxies and search coverage \cite{staufer2025audit} used to validate the failure, or standardized auditing, mapping the challenge to specific regulatory or ethical benchmarks~\cite{costanza2022audits,birhane2024ai}. Assurance 2.0~\cite{bloomfield2024defeaters, rushby2015understanding} argues that a challenge without an explicit rationale \textbf{\textit{(R6)}} and underlying premises and assumptions \textbf{\textit{(R7)}} cannot be meaningfully evaluated or rebutted and hence, is simply unverifiable. Similarly, it requires documentation of severity and implications \textbf{\textit{(R8)}} to contextualize why a defeater was considered significant and to avoid defeater gaming. Next, the operational context and boundaries to which the defeater applies \textbf{\textit{(R9)}} are equally critical. Rushby~\cite{rushby2015understanding} argues that a challenge without a defined scope is not falsifiable, a position consistent with operational design domain frameworks in autonomous systems~\cite{bloomfield2021safety}. Finally, identifying the affected component \textbf{\textit{(R10)}} is essential for localized, targeted mitigation. This prevents scoping errors, where a failure is either ignored or incorrectly over-generalized, ensuring the defeater is evaluated only within its valid technical boundaries.

\subsubsection{Mitigation Rigor \& Robustness}In defeater analysis, documenting the rigor of mitigation is critical: evidence for resolving a defeater must clearly indicate how the mitigation is complete and robust. Safety case guidance~\cite{ACWG21, hobbs2024driving} stresses that mitigation strategies should include explicit boundary conditions and limitations \textit{(\textbf{R11})}, particularly in modern AI-driven systems where mitigation approaches are often probabilistic or algorithmic~\cite{feffer2024red}. To support safety assurance case maintenance, monitoring arrangements and re-evaluation triggers \textit{(\textbf{R12})} (if available) should be included, as systems and contexts evolve~\cite{ ai2023artificial}. Next, regulatory standards and frameworks often require that residual risks \textit{(\textbf{R13})} be recorded. Finally, a meta-evaluation of the mitigation is helpful for addressing the growing concerns about the lack of detail regarding its reliability 
and validity \textit{(\textbf{R14})}.

\subsubsection{Dialectic Reasoning} This is a central theme in Assurance 2.0~\cite{bloomfield2020assurance} and involves evaluating claims through structured consideration of opposing perspectives~\cite{walton1998new}. Multiple studies have called for its inclusion in assurance documentation~\cite{assurance2021assurance, bloomfield2020assurance}, yet fragmented practices across notations (e.g., GSN, EA~\cite{kelly2004goal,goodenough2015eliminative}) often prevent systematic adoption. Critically, much of the insight from earlier reviews is lost if challenge-response threads, rebuttals, and counterfactual scenarios \textit{(\textbf{R15})} are not preserved~\cite{ACWG21}. Rejected arguments can be as informative as accepted ones, providing essential audit evidence that the final artifact alone cannot convey. Equally important is tracking resolution status and decision-making across review iterations \textit{(\textbf{R16})}~\cite{hobbs2024driving}, ensuring that the evolution of the argument remains visible and contestable. Our 5W1H-based framework provides a structured method that directly enables practitioners to critically reflect on defeaters. 

\section{Defeater Cards: Representation and Structure}
\label{sec:card}

In this section, we present the six dimensions of our framework for defeater analysis and documentation. Like prior efforts \cite{mitchell2019model,gebru2021datasheets}, they are not exhaustive; rather, they are designed to guide critical reflection and capture relevant details for stakeholders across the defeater lifecycle.

\begin{figure}[t]
\input{card}
\caption{Proposed \emph{Defeater Card} template with sections based on 5Ws, 1H framework to support dialectic reasoning. }
\end{figure}

\subsection{Metadata} 

The metadata section captures basic information such as unique identifiers, providing foundational context for versioning, traceability, and cross-referencing within and across systems. Most fields can be auto-filled via tool support, reducing documentation overhead and errors.

\noindent \textbf{Defeater ID:} A unique identifier assigned to each defeater for tracking and reference. 

\noindent \textbf{Defeater Tag:} Keywords or labels categorizing the defeater (e.g., data bias, invalid assumption) to facilitate filtering and search, typically internal to organizations.

\noindent \textbf{Defeater Type:} The high-level classification of the defeater, such as \textit{logical, evidence-based, or contextual}, based on established taxonomies \cite{gohar2025taxonomy,greenwell2006taxonomy}. This is particularly useful for stakeholders as it supports a systematic approach to defeaters.  

\noindent \textbf{Version:} Which version of the defeater is this? And how has it evolved? This is critical for reviewers to understand the lifecycle of defeaters \cite{assurance2021assurance}.

\noindent \textbf{Date:} When the defeater was identified, allowing stakeholders to assess whether original assumptions remain valid as capabilities and environments evolve \cite{mitchell2019model}.

\noindent \textbf{Status:} The current state of the defeater (e.g., open, mitigated, residual), as not all can be mitigated \cite{goodenough2015eliminative}.

\subsection{The What? (Identification and Evaluation)}

The first question focuses on \textit{what} is being evaluated, identifying which part of the assurance case is being challenged and providing the necessary context.

\noindent \textbf{Affected Node(s):} This captures the assurance case node impacted by the defeater. A single defeater can affect multiple nodes and trigger secondary defeaters elsewhere in the argument~\cite{bloomfield2024defeaters}, revealing hidden dependencies and interaction chains, and serving as a proxy for severity (see \textit{Why?} dimension). In practice, practitioners may begin with the directly challenged node and expand iteratively; automated tools \cite{Yu25,AISupported} can assist in identifying non-obvious propagation paths.

\noindent \textbf{Defeater Description:} A concise explanation of the defeater, summarizing the potential issue, failure, or counterargument it represents. 

\noindent \textbf{Source:} What is the source of the defeater? For example, was the defeater identified via expert review, incident reporting (e.g., Drones incident database, AI incident database), or tooling? Documenting the source allows reviewers to assess the challenge's credibility and limitations as different sources have distinct biases, coverage gaps, and reliability profiles that affect defeater validity \cite{ACWG21}. Understanding the source helps evaluate whether the analysis relied on diverse perspectives or systematic methods, and supports identification of broader patterns across systems \cite{bloomfield2021safety}. This also helps distinguish between empirically grounded and hypothetical defeaters.

\subsection{The Why? (Justification and Rationale)} 

Assurance case evaluation is often undermined by superficial arguments and "confidence-inflating" defeaters \cite{gohar2025taxonomy,graydon2017investigation}, a trend mirrored in ML research \cite{head2015extent,ganesh2025different}. Even quantitative confidence measures can be manipulated to create an illusion of rigor \cite{graydon2017investigation}. To counteract this, defeaters should be accompanied by well-justified rationales, and details on severity, impact, and inclusion criteria \cite{assurance2021assurance,ACWG21,bloomfield2021safety}. Such transparency is especially critical when a defeater is dismissed, as it allows reviewers to audit the underlying reasoning.

\noindent \textbf{Rationale:} This field captures the explicit rationale and justification for including (or later dismissing) a defeater. Examples include unsupported inference steps, known limitations of the evaluation method, or mismatches between the evidence scope and the claim.

\noindent \textbf{Underlying Assumption:} Unlike system-level assumptions, these are analytical or contextual premises used to interpret evidence and assess defeater plausibility. They encompass beliefs regarding data representativeness, evidence reliability, and operational conditions \cite{mitchell2019model}. Explicitly documenting these assumptions ensures reasoning remains transparent, allowing reviewers to trigger a reassessment if the underlying context shifts.

\noindent \textbf{Severity/Potential Impact:} Each defeater is characterized by its potential risk to the argument. \textit{Severity} measures the degree of challenge to the reasoning (e.g., a minor gap vs. a critical failure), while \textit{Impact} assesses consequences such as failure propagation or increased uncertainty. \textit{Likelihood} distinguishes between theoretical concerns and probable weaknesses. Together, these provide a multidimensional profile for systematic evaluation \cite{bloomfield2021safety}, which can be captured via severity-likelihood matrices, scenario-based scoring \cite{novelli2024ai,koessler2023risk} or domain-specific scales \cite{dezfuli2011nasa}.

\subsection{The Who? (Stakeholders, Tools, and Expertise)} To support audit and review, defeater cards should document who participated in the defeater process, including humans (individuals or teams) and automated methods (tools, LLM judges, etc.)~\cite{ACWG21,staufer2025audit,costanza2022audits}

\noindent\textbf{Human (Individual and Team):} Documenting their domain expertise and experience enables reviewers to assess coverage, identify blind spots, and evaluate the reliability of their reasoning. In complex or human-AI systems, diverse perspectives are essential to surfacing edge cases and unusual scenarios \cite{feffer2024red}. To maintain privacy, this should focus strictly on professional background and expertise rather than personal identifiers.

\noindent\textbf{Automated:} Identification increasingly relies on automated methods such as semantic analysis, formal methods, and LLM frameworks \cite{muram2023attest, gohar2024codefeater,Yu25}. However, every method has inherent limitations, such as training biases in LLMs \cite{wang2024large} or the specification dependency of formal methods, which can affect the credibility of the analysis \cite{Graydon25}. Disclosing the specific tools used ensures transparency and provides reviewers with the necessary context to identify potential gaps and limitations.

\subsection{The When? (Temporal and Lifecycle Relevance)} Defeaters manifest at different stages with varying temporal characteristics, from immediate risks to latent vulnerabilities triggered by specific conditions. This dimension maps "when" a defeater emerges, and is essential for designing monitoring strategies and maintaining assurance as systems evolve \cite{denney2015dynamic}.

\noindent \textbf{Event/Trigger:} Event/Trigger: This field identifies conditions that activate a defeater, such as sensor drift, software updates, or environmental shifts. It also captures speculative or future-dated defeaters, such as anticipated regulatory changes. Current frameworks typically lack mechanisms
for revisiting such speculative defeaters as conditions change~\cite{maksimov2019survey}. Documenting both concrete triggers and latent conditions enables runtime monitoring, 
distinguishes between immediate and future risks, and ensures that speculative concerns are revisited as operational contexts shift.

\noindent \textbf{Defeater Frequency:} This captures whether a defeater is one-time, periodic, continuous, or event-driven. Frequency affects resource allocation: high-frequency threats may require automated monitoring or streamlined review. In socio-technical contexts, frequency should be recorded alongside relevant observation windows.

\noindent \textbf{Phase/Lifecycle Stage:} Specifies when a defeater is relevant (e.g., design vs. operation). This context determines the type of evidence available to address the challenge and identifies necessary regulatory checkpoints. Explicitly mapping lifecycle stages potentially reveals cross-phase dependencies, e.g., design assumptions that may transition into operational monitoring requirements, ensuring the safety case remains valid throughout the system's life, not just at a single point in time.

\subsection{The Where? (Context and Scope)} This section captures the context in which a defeater emerges along two axes: the specific system component affected and the operational context and scope in which it emerges. By enforcing explicit scoping, the framework avoids selective reporting and supports comprehensive coverage analysis \cite{gohar2025taxonomy}.

\noindent \textbf{System/Component:} Identifies the part of the system (hardware, software module, subsystem, or process) where the defeater is relevant, ensuring reviewers understand its locus of impact. Documenting this context localizes impact, accounts for context dependence, and reveals coverage gaps where critical components or scenarios remain unexamined \cite{bloomfield2021safety,hawkins2015weaving}.

\noindent \textbf{Operational Context:} Many modern safety-critical system, such as autonomous vehicles,  operate across heterogeneous and dynamic environments.  This field specifies the environmental, configuration, or usage conditions under which the defeater occurs, clarifying scope and applicability. This is often relevant for runtime adaptive systems. 

\subsection{The How? (Mitigation and Response)} 

This dimension captures how defeaters are monitored, mitigated, or otherwise managed. Current frameworks often overlook (1) the limitations of proposed mitigations and (2) their reliability and validity~\cite{staufer2025audit}. Explicit documentation of these considerations captures developers' insights into current or potential future remedies and helps reviewers assess their adequacy and rigor.

\noindent \textbf{Monitoring and Mitigation:} This field primarily records the mitigation and monitoring strategies to handle the defeater. This can include discussion of computational methods, making changes to the design of the system, making changes to the operation of the system, e.g., by limiting the conditions under which the system is used, making changes to the assurance argument, e.g., adding an independent source of evidence or generating additional evidence for the confidence argument. In instances where mitigation is infeasible, e.g., due to epistemic constraints or conditional risks during runtime or residual risks, monitoring is required to flag violations and then follow up with relevant procedures~\cite{denney2015dynamic}. 

\noindent \textbf{Reliability and Validity:} This field elicits from the defeater card user any analytic or experiential information or regarding the likelihood that the intended mitigation will perform failure-free as expected.  It also encourages reasoning about whether the indicated mitigation for this defeater will perform as required, and how that might be confirmed or demonstrated.

\noindent \textbf{Limitations (e.g., Sociotechnical, disagreements, project constraints)} This field encourages the stakeholders to explicitly consider and report limitations of the proposed approach(es). These might also include disagreements between reviewers and/or the team, which are important signals in a dialectic framework \cite{bloomfield2020assurance}. 

\noindent \textbf{Residual Risks:} Not all defeaters can be mitigated completely, in which case residual risks remain. Acknowledging and highlighting these supports transparency, can prevent unforeseen failures, and enables appropriate monitoring plans to be put in place.

\subsection{Additional comments} This field allows free-form discussion of additional concerns by the different stakeholders involved.

\section{RQ2: Case Studies}
\label{sec:applications}

\begin{figure*}[!t]
\definecolor{CardFrame}{HTML}{404040} 
\definecolor{CardBackground}{HTML}{FFFFFF}
\definecolor{SeparatorGray}{HTML}{D0D0D0}



\newcommand{\cardSection}[3]{ 
    \vspace{2pt}\noindent
    \parbox{\textwidth}{
        \textbf{#1} (\small#2) \textbar\ #3
    }
}
\noindent
\begin{minipage}{\textwidth} 
\begin{tcolorbox}[
    colframe=CardFrame,
    colback=CardBackground,
    boxrule=0.4pt,
    boxsep=1.5pt, 
    left=1mm, right=2mm, top=2mm, bottom=2mm, 
    before skip=5pt,
    after skip=5pt,
    width=\textwidth
]

    {\centering\textbf{\normalsize{Defeater Card}}\par} 
    \small
    \textbf{Metadata}
    \vspace{0.8mm}

\setlength{\tabcolsep}{2pt}
\begin{tabular}{@{}p{2.3cm}p{3.8cm}p{3.5cm}p{1.6cm}p{1.8cm}p{2cm}@{}}
\textbf{Defeater ID} & \textbf{Defeater Tag} & \textbf{Defeater Type:} & \textbf{Version: } & \textbf{Date:} & \textbf{Status:} \\
BVLOS-DF-01 & Sensor Fusion Confidence & Evidence Validity  & 1.0 & March 2025 & Open \\
\end{tabular}

    {\color{SeparatorGray}\hrule} 

    \cardSection{What?}{Identification and Evaluation}{\textit{What is being evaluated? (e.g., what argument, evidence, etc.)}}
    \vspace{-0.7em}
    \begin{itemize}[leftmargin=4mm, nosep, itemsep=0pt, before=\vspace{0pt}, after=\vspace{2pt}]
        \item \textbf{Affected Node:} \textit{Claim C3.2 – "The Detect-and-Avoid (DAA) subsystem reliably identifies and classifies all nearby airborne hazards during BVLOS flight"}
        \item \textbf{Defeater Description:} \textit{During BVLOS operations, the DAA system fuses inputs from ADS-B, radar, and optical sensors. Under certain conditions (e.g., partial occlusion, or radar clutter), sensor disagreement lowers fusion confidence, causing the system to alternate between “avoid” and “ignore,” leading to false or missed obstacle detections.}
        
        \item \textbf{Source:} Detected in post-flight analysis of simulated airspace trials and confirmed in real-flight telemetry during BVLOS corridor testing (Arizona desert, 2024).
    \end{itemize}

    {\color{SeparatorGray}\hrule} 

    \cardSection{Why?}{Justification and Rationale}{\textit{Why is this important to the validity or credibility of the argument or evidence?}}
    \vspace{-0.6em}
    \begin{itemize}[leftmargin=4mm, nosep, itemsep=0pt, before=\vspace{0pt}, after=\vspace{2pt}]
        \item \textbf{Rationale:} The assurance argument presumes that multi-sensor fusion provides redundancy and thus improves reliability. However, when the calibration of individual sensors is compromised, fusion amplifies correlated errors. This defeater exposes an unacknowledged vulnerability: the confidence metric itself becomes an unreliable proxy for detection quality.
        \item \textbf{Underlying Assumptions:} Sensor noise distributions are stationary across environments.
        \item \textbf{Severity/Potential Impact: } Moderate-to-High. In two out of eight test flights, confidence dips led to delayed evasive maneuvers (>1.5 s latency), violating the 2-s detection-to-action requirement for BVLOS safety envelopes 
    \end{itemize}
    
    {\color{SeparatorGray}\hrule} 

    \cardSection{Who?}{Stakeholders, Expertise \& Contributors}{\textit{Who are the auditors/reviewers? (e.g., expertise coverage, teams etc.)}}
    \vspace{-0.6em}
    \begin{itemize}[leftmargin=4mm, nosep, itemsep=0pt, before=\vspace{0pt}, after=\vspace{2pt}]
        \item \textbf{Automated:} N/A
        \item \textbf{Human:} \textbf{Expertise/Background:} Safety assurance, ML/AI Perception, 
        BVLOS regulatory compliance.
    \end{itemize}
    
    {\color{SeparatorGray}\hrule} 

    \cardSection{When?}{Temporal \& Lifecycle Relevance}{\textit{When does the defeater emerge? (e.g., under stress conditions, etc.)}}
    \vspace{-0.6em}
    \begin{itemize}[leftmargin=4mm, nosep, itemsep=0pt, before=\vspace{0pt}, after=\vspace{2pt}]
        \item \textbf{Event/Trigger: }Occurs when operating in mixed-visibility environments e.g., near-horizon glare or partial cloud cover.
        \item \textbf{Defeater Frequency:} Appeared in 15\% of BVLOS test missions (n = 100).
        \item \textbf{Phase:} Emerged in Operational Testing after certification prototype.
    \end{itemize}
    
    {\color{SeparatorGray}\hrule} 
    
    \cardSection{Where?}{Context and Scope}{\textit{Where within the system or operational context does this defeater apply or manifest?}}
    \vspace{-0.6em}
    \begin{itemize}[leftmargin=4mm, nosep, itemsep=0pt, before=\vspace{0pt}, after=\vspace{2pt}]
        \item \textbf{System/Component:} Detect-and-Avoid subsystem, Multi-Sensor Fusion Layer, Confidence Estimator module.
        \item \textbf{Operational Context:} Daytime BVLOS operations in mixed-visibility environments  
    \end{itemize}
    
    {\color{SeparatorGray}\hrule} 

    \cardSection{How?}{Mitigation \& Response}{\textit{How is this defeater detected, monitored, or mitigated?}} 
    \vspace{-0.6em}
    \begin{itemize}[leftmargin=4mm, nosep, itemsep=0pt, before=\vspace{0pt}, after=\vspace{3pt}]
        \item \textbf{Monitoring \& Mitigation:} Add soft angular limits to the gimbal control system to prevent optical sensor alignment within critical solar angles (calculated pre-flight)
        \item \textbf{Reliability \& Validity:} Statistical variance analysis across repeated flights (N = 50) under changing light conditions remained within 95\% confidence bands. Cross-validation with manual labeling achieved Cohen’s K = 0.82 (strong agreement) for true conflicts.
        \item \textbf{Limitations:} Not available on all airframes
        \item \textbf{Residual Risks:} Residual missed-intruder risk remains and must be handled by conservative alerting and contingency procedures (e.g., quick descent, landing in place if possible). Multiple sensor degradation still a risk.
    \end{itemize}

\end{tcolorbox}
\end{minipage}
\caption{Example Defeater Card for an sUAS Assurance Case.}
\label{fig:DC1}
\end{figure*}

We now present two case studies that apply the Defeater Cards framework to diverse safety-critical domains. We chose two very different domains that have appeared in the safety case literature for the purpose of generality. These examples demonstrate the practical applicability of Defeater Cards and their support for transparent, standardized, and rigorous documentation of defeaters for safety case evaluation. The first is for an sUAS scenario, and the second is for an embedded nano-device that uses molecular programming. We have also developed and made available an open-source repository with twelve additional defeater cards from seven safety-critical systems in diverse domains as a baseline for researchers and practitioners.

\subsection{Small Uncrewed Aerial Systems (sUAS)}

For our first example, we use the defeater \textit{D1} from the sUAS assurance case in Figure~\ref{fig:AC}, which challenges the claim \textit{"G1.2: The Detect-and-Avoid (DAA) subsystem reliably identifies all nearby airborne hazards during BVLOS flight."} This is presented in Figure~\ref{fig:DC1}.

In contrast to the original defeater, we can see the contextual details that our framework provides. The \textit{Why?} section reveals that the assurance argument assumes multi-sensor fusion provides redundancy, yet under sensor miscalibration, fusion amplifies correlated errors rather than suppressing them, undermining the very claim it was meant to support. The \textit{How?} section documents not only the mitigation but its limitations: mitigation is unavailable on all airframes, and residual missed-intruder risk remains unresolved. Practitioners could, and have (e.g., GSN 3.0 \cite{GSNstandard}), extended the assurance case itself to capture such detail, but prior work has shown that this quickly becomes unwieldy and degrades readability \cite{bloomfield2021safety,rushby2015understanding}. Moreover, our framework provides sufficient structure to support critical reflection in the first place. Our artifact contains additional defeater cards that apply to other parts of the sUAS system, e.g., configurations, battery, etc.

\subsection{Molecular Programming}

Our second example, presented in Figure~\ref{fig:DC2}, provides a defeater card for a partial assurance case of a molecular program, \cite{Lapteva22}, with safety-critical applications such as targeted drug delivery \cite{Lutz22}. In a molecular program, computational logic is encoded directly into molecules. As such, defeaters often arise from unmodeled biochemical interactions, inaccurate assumptions about molecular kinetics, or simulation-to-wet-lab discrepancies \cite{Tun15}. 

This case study highlights several observations. First, it demonstrates that the defeater card framework extends naturally to emerging computational domains, beyond software-intensive systems. Second, the \textit{Why?} and \textit{Where?} sections surface assumptions about molecular kinetics that were implicit in the original assurance argument but not explicitly recorded, underscoring how our framework functions as an elicitation tool, as well as a documentation artifact. More broadly, this example illustrates that defeater cards can surface gaps earlier in the lifecycle, before deployment, especially important in safety-critical domains~\cite{Leveson95}.

\begin{figure*}[!ht]
\definecolor{CardFrame}{HTML}{404040} 
\definecolor{CardBackground}{HTML}{FFFFFF}
\definecolor{SeparatorGray}{HTML}{D0D0D0}



\newcommand{\cardSection}[3]{ 
    \vspace{2pt}\noindent
    \parbox{\textwidth}{
        \textbf{#1} (\small#2) \textbar\ #3
    }
}
\noindent
\begin{minipage}{\textwidth} 
\begin{tcolorbox}[
    colframe=CardFrame,
    colback=CardBackground,
    boxrule=0.4pt,
    boxsep=1.5pt, 
    left=1mm, right=2mm, top=2mm, bottom=2mm, 
    before skip=5pt,
    after skip=5pt,
    width=\textwidth
]

    {\centering\textbf{\normalsize{Defeater Card}}\par} 
    \vspace{3pt} 
    \vspace{2pt} 
    \small
  
    \textbf{Metadata}
    \vspace{0.8mm}

\setlength{\tabcolsep}{2pt}
\begin{tabular}{@{}p{2.3cm}p{3.8cm}p{3.5cm}p{1.6cm}p{1.8cm}p{2cm}@{}}
\textbf{Defeater ID} & \textbf{Defeater Tag} & \textbf{Defeater Type:} & \textbf{Version: } & \textbf{Date:} & \textbf{Status:} \\
LC-02 & DNA Temporal Logic  & Logical  & 3.0 & Aug 2025 & Mitigated \\
\end{tabular}

    {\color{SeparatorGray}\hrule} 

    \cardSection{What?}{Identification and Evaluation}{\textit{What is being evaluated? (e.g., what argument, evidence, etc.)}} 

    \vspace{-2mm}
    
    \begin{itemize}[leftmargin=4mm, nosep, itemsep=0pt, before=\vspace{0pt}, after=\vspace{2pt}]
        \item \textbf{Affected Node:} G1.1 – "The molecular circuit reliably encodes input event order within specified kinetic tolerances"
        \item \textbf{Defeater Description: } The argument claims that DNA strand-displacement temporal logic circuits (TL-circuits) maintain reliable ordering of inputs under varying reaction intervals. However, 
        deviations in toehold kinetics and input concentration shift output timing, causing temporal misalignment between expected and actual reaction order. Description based on \cite{Lapteva22}.
        \item \textbf{Source:} Drift observed during laboratory experiments. 
    \end{itemize}

    {\color{SeparatorGray}\hrule} 

    \cardSection{Why?}{Justification and Rationale}{\textit{Why is this important to the validity or credibility of the argument or evidence?}} 

    \vspace{-2mm}
    
    \begin{itemize}[leftmargin=4mm, nosep, itemsep=0pt, before=\vspace{0pt}, after=\vspace{2pt}]
        \item \textbf{Rationale:} The kinetic tolerances specified were derived from idealized conditions that do not account for concentration variability or toehold sensitivity observed. 
        \item \textbf{Underlying Assumptions:} Assumes that toehold shortening uniformly improves robustness without side-effects on false activation rates.
        \item \textbf{Severity/Potential Impact:} High. Crosstalk directly violates the circuit's temporal discrimination function. 
    \end{itemize}
    
    {\color{SeparatorGray}\hrule} 

    \cardSection{Who?}{Stakeholders, Expertise \& Contributors}{\textit{Who are the auditors/reviewers? (e.g., expertise coverage, teams etc.)}}

    \vspace{-2mm}

    \begin{itemize}[leftmargin=4mm, nosep, itemsep=0pt, before=\vspace{0pt}, after=\vspace{2pt}]
        \item \textbf{Automated:} Visual DSD simulator \& MATLAB post-processing detected kinetic divergence.
        \item \textbf{Human:}  Molecular computing researcher (circuit design), bioengineer (experimental validation),

    \end{itemize}
    
    {\color{SeparatorGray}\hrule} 

    \cardSection{When?}{Temporal \& Lifecycle Relevance}{\textit{When does the defeater emerge? (e.g., under stress conditions, etc.)}}

    \vspace{-2mm}

    \begin{itemize}[leftmargin=4mm, nosep, itemsep=0pt, before=\vspace{0pt}, after=\vspace{2pt}]
        \item \textbf{Event/Trigger:} Observed during 
        phase two validation of two-input AND/OR circuits when input delay exceeded 20 min
        \item \textbf{Defeater Frequency:} Observed consistently with longer toeholds (6–7 nt) across three independent runs
        \item \textbf{Phase/Lifecycle:} Verification and replication phase of molecular computing testbed
    \end{itemize}
    
    {\color{SeparatorGray}\hrule} 
    
    \cardSection{Where?}{Context and Scope}{\textit{Where within the system or operational context does this defeater apply or manifest?}}

    \vspace{-2mm}

    \begin{itemize}[leftmargin=4mm, nosep, itemsep=0pt, before=\vspace{0pt}, after=\vspace{2pt}]
        \item \textbf{System/Component:} Logic gate strands
        \item \textbf{Operational Context:} DNA self-assembly 
    \end{itemize}
    
    {\color{SeparatorGray}\hrule} 

    \cardSection{How?}{Mitigation \& Response}{\textit{How is this defeater detected, monitored, or mitigated?}} 
    
    \vspace{-2mm}
    
    \begin{itemize}[leftmargin=4mm, nosep, itemsep=0pt, before=\vspace{0pt}, after=\vspace{3pt}]
        \item \textbf{Monitoring \& Mitigation:} Used dual-channel fluorescence tracking to detect kinetic lag between logic modules. Optimized toehold lengths to 5–7 nt and normalized input concentrations (100–300 nM) to stabilize reaction rates.
        \item \textbf{Reliability and Validity:} Re-evaluation at 10–30 min input intervals yielded consistent fluorescence peaks within ± 5\% of predicted timing, demonstrating reproducibility under controlled conditions.
        \item \textbf{Limitations:} Mitigation does not address ionic-strength variability or long-term storage effects;  re-calibration required for field-deployable settings.
        \item \textbf{Residual Risks:} Reaction fidelity remains sensitive to ambient temperature and reagent degradation; cumulative timing errors may compound in cascaded circuits; fluorescence readout may introduce bias in kinetic analysis.
    \end{itemize}
    
    {\color{SeparatorGray}\hrule} 

    \vspace{1mm}
    \textbf{Additional Comments:}
    \begin{itemize}[leftmargin=4mm, nosep, itemsep=0pt, before=\vspace{0pt}, after=\vspace{2pt}]
        \item \textbf{Notes:} See \cite{Lapteva22} and its Supporting Information.  
    \end{itemize}

\end{tcolorbox}
\end{minipage}
\caption{Example Defeater Card for a Molecular Program \cite{Lapteva22} Assurance Case.}
\label{fig:DC2}
\end{figure*}

\section{Lessons learned}
\label{sec:discussion}

We describe below some lessons learned from our experience of developing, using, and reviewing defeater cards. We focus on those lessons that may be useful to projects wanting to use defeater cards and on lessons that point the way forward.   

\noindent \textbf{Toward better understanding of defeaters}.  Developing the defeater card reveals limits to a developer's knowledge of the operational context.  We noted that several defeaters in the literature are too high-level to be useful, basically of the form “Unless it doesn't work."  The defeater card {5W+1H} format encourages users to more thoroughly reason about and report how that breakage might happen, what effects it might have, and how much priority should be given to preparing for its occurrence. 

\noindent \textbf{Toward improved maintenance of safety assurance cases in dynamic scenarios.}  The defeater cards’ organized clarity is especially beneficial when a person other than the originator needs to update the card, as when a potential defeater later is operationalized, triggering the need for a new safety requirement (often foreseen in the card’s mitigation field). We thus recommend, as an initial use case for defeater cards, those dynamic safety assurance cases where developers and maintainers are separate teams. Further, defeater cards can reduce the impact of personnel turnover-induced knowledge loss~\cite{aghajani2020software}. Additionally, because defeater cards describe potential future risks to satisfying the safety requirements, they help position a project to anticipate and adapt to likely changes.  

\noindent \textbf{Enabling better reviews and audits.}   Reviews and audits of safety assurance cases can benefit from defeater cards’ scaffolding of needed information.  We found that the defeater cards’ structure and guidance resulted in users writing cards that reviewers could read and understand. Contextual explanations and assumptions may be unknown or non-intuitive to the reviewer or auditor, and making these explicit both facilitates and enhances evaluation of the safety case. 

\noindent \textbf{Balancing practicality and completeness.} Structured documentation improves transparency, consistency, and traceability \cite{aghajani2020software}. However, achieving comprehensive coverage requires significant time and resources, creating an inherent trade-off between completeness and feasibility. The appropriate balance will vary across domains, regulatory contexts, and organizational settings. Accordingly, defeater cards prioritize documenting critical contextual information without aiming for exhaustiveness, while remaining general enough for cross-domain use. Future study is needed to examine how stakeholders will interpret and apply defeater cards in practice.

\noindent \textbf{Residual risk of defeater hacking in practice.} Despite the structured approach introduced by defeater cards, the risk of defeater hacking remains, i.e., strategically omitting, misclassifying, or selectively including defeaters to create an illusion of robustness \cite{graydon2017investigation}. Practioners may focus on low-impact issues while overlooking systemic concerns, reducing defeater analysis to a compliance exercise \cite{pasman2018trustworthy,cave2006independent}. 
Addressing this could involve formal completeness criteria, automated elicitation from logs and failure histories \cite{denney2015dynamic}, and third-party assessments. 

\noindent \textbf{Toward structured documentation to report defeaters.}  The proposed framework provides practitioners and safety analysts with a structured approach to systematically document, analyze, and mitigate defeaters while supporting auditability. Contextual explanations and assumptions may be unknown or non-intuitive to the reviewer or auditor, and making these explicit supports evaluation of the safety case. As well, use of defeater cards may enhance LLM-based approaches~\cite{gohar2024codefeater, AISupported, Graydon25,Yu25} by providing the contextual information that they require for automated evaluation of safety assurance cases.

\section{Conclusion}
\label{sec:conclusion}

We introduce Defeater Cards, a structured framework for documenting defeaters in safety assurance cases. Through a systematic review of the literature and our experience developing and analyzing defeaters, we identified key challenges that informed the design of our framework. Defeater Cards are designed for use during assurance case development to standardize documentation of defeater reasoning. They proactively help capture and characterize potential obstacles to the soundness of the safety arguments.  By structuring defeater reasoning, Defeater Cards improve transparency, support updates and auditing, and strengthen the validity and rigor of safety arguments. We demonstrate applicability through two cross-domain case studies. We also release a public repository with templates and example cards across seven systems. We envision Defeater Cards as living artifacts that evolve through community adoption, domain-specific extensions, and empirical validation.

\section*{Acknowledgments}
This work was funded by grant 80NSSC23M0058 from the National
Aeronautics and Space Administration (NASA) and by NSF CCF 1900716 and CCF-2211589.

{\small
\bibliographystyle{IEEEtran}
\bibliography{sample-base}}

\end{document}